\documentstyle[12pt]{article}

\newcommand{\be}{\begin{equation}}
\newcommand{\ee}{\end{equation}}
\newcommand{\bea}{\begin{eqnarray}}
\newcommand{\eea}{\end{eqnarray}}
\newcommand{\gsim}{ \mathop{}_{\textstyle \sim}^{\textstyle >} }
\newcommand{\lsim}{ \mathop{}_{\textstyle \sim}^{\textstyle <} }

\def\lm{\lambda}

\def\Sig{\Sigma}

\def\pdy{\partial_y}
\def\vev#1{\langle #1 \rangle}

\newcommand\bh{\bar{H}}
\newcommand\hc{H^c}
\newcommand\bhc{\bar{H}^c}

\newcommand\bphi{\bar{\phi}}
\newcommand\phic{\phi^c}
\newcommand\bphic{\bar{\phi}^c}

\def\yq3{y_{Q_3}}
\def\yq2{y_{Q_2}}
\def\yq1{y_{Q_1}}
\def\yu3{y_{U^c_3}}
\def\yu2{y_{U^c_2}}
\def\yu1{y_{U^c_1}}
\def\yd3{y_{D^c_3}}
\def\yd2{y_{D^c_2}}
\def\yd1{y_{D^c_1}}
\def\yl3{y_{L_3}}
\def\yl2{y_{L_2}}
\def\yl1{y_{L_1}}
\def\ye3{y_{E^c_3}}
\def\ye2{y_{E^c_2}}
\def\ye1{y_{E^c_1}}

\def\bx{\bar{X}}
\def\bm{\bar{M}}
\def\tr{{\rm tr}}
\def\barx{\bar{x}}
\def\barm{\bar{m}}
\def\dx{x-\bar{x}}
\def\dm{m-\bar{m}}

\begin{document}

\begin{titlepage}
\null
\begin{flushright}
{\tt hep-ph/0108002}\\
UT-954\\
August  2001
\end{flushright}
\vskip 1cm
\begin{center}
{\Large\bf Doublet-Triplet Splitting and Fat Branes}

\lineskip .75em
\vskip 1.5cm

\normalsize

 {\large Nobuhito Maru$^{a}$}
\footnote{e-mail address: maru@hep-th.phys.s.u-tokyo.ac.jp; 
JSPS Research Fellow.} 

\vspace{1cm}

{\it $^{a}$Department of Physics, University of Tokyo, \\
Tokyo 113-0033, JAPAN} \\


%
\vspace{18mm}

{\bf Abstract}\\[5mm]
{\parbox{13cm}{\hspace{5mm}
%

We consider the doublet-triplet splitting problem 
in supersymmetric SU(5) grand unified theory in five dimensions 
where the fifth dimension is non-compact. 
We point out that an unnatural fine-tuning of parameters 
in order to obtain the light Higgs doublets is not required 
due to the exponential suppression of the overlap of the wave functions. 

}}

\end{center}

\end{titlepage}


The grand unified theory (GUT) \cite{GUT} 
is one of the most elegant scenarios 
in particle physics because of its aesthetic point of view and 
various interesting physical features. 
In particular, 
the success of the gauge coupling unification with 
the minimal supersymmetric standard model (MSSM) \cite{unification} 
motivates us to consider the supersymmetric (SUSY) GUT seriously. 
In GUT, 
there is a serious problem called 
``the doublet-triplet splitting problem". 
After GUT symmetry breaking, 
the Higgs triplets and the Higgs doublets in general obtain 
the GUT scale mass because these belong to the same multiplet. 
This requires an unnatural fine-tuning of parameters 
to obtain the light Higgs doublets \cite{SUSYGUT}. 
In the minimal SU(5) GUT case, 
the mass term $MH\bar{H}$ is tuned for the coupling $\bar{H} \Sigma H$ 
($\Sigma$: the adjoint under SU(5)) to obtain the Higgs doublet 
with weak scale mass, while leaving the Higgs triplets 
with GUT scale mass to avoid a fast proton decay 
by dimension five operators \cite{SYW}. 
Many people have tried to solve this problem from 
various points of view 
\cite{Missing}-\cite{deconst}.

Arkani-Hamed and Schmaltz \cite{AS} have proposed 
an interesting mechanism to generate exponentially small coupling 
in the context of extra dimensions. 
They discussed that the hierarchies of Yukawa couplings 
can be explained by the slight displacement of the standard model 
field wave functions inside four dimensional domain wall 
in higher dimensional space-time. 
Even if there is a parameter of order one in the fundamental theory, 
it is highly suppressed in the effective theory due to 
the small overlap of wave functions.

In this letter, 
we apply this mechanism to the doublet-triplet splitting problem. 
It is pointed out that an unnatural fine-tuning of parameters 
to obtain the light Higgs doublets is not required in this scenario, 
{\it i.e.} at most the tuning of ${\cal O}(1)$ orders of magnitude. 
For simplicity, 
we consider the SUSY GUT in five dimensions 
where the fifth dimension is non-compact. 
The action of the Higgs sector is 
\bea
\label{Higgsaction}
S &=& \int d^4xdy \left[ \int d^4 \theta 
(H^{\dagger} e^{-V} H + H^{c{\dagger}} e^V \hc 
+ \bh^{\dagger} e^V \bh 
+ \bar{H}^{c{\dagger}} e^{-V} \bhc ) \right. \nonumber \\ 
&+& \left\{ \int d^2 \theta 
\left( {\hc ( \pdy + X(y) + M) H 
+ \bhc ( \pdy + \bx(y) + \bm) \bh } \right) \right. \\ 
&+& \delta(y) 
\int d^2 \theta \left. \left.
\left( \lm_1 {\rm tr}(X^2 \Sig) 
+ \lm_2 {\rm tr}(\bx^2 \Sig) 
+ \lm_3 {\rm tr}(X \Sig^2) 
+ \lm_4 {\rm tr}(\bx \Sig^2) 
+ \frac{1}{2} m_0 {\rm tr}(\Sig^2) \right) 
+ {\rm h.c.} \right\} \right], \nonumber
\label{action}
\eea
where $H (\bh), \hc (\bhc)$ are left-handed 
(charge conjugated right-handed) chiral 
${\cal N} = 1$ in four dimensional 
superfield components of the single ${\cal N} = 1$ 
in five dimensional chiral superfield 
$H({\bf 5}) = \left( H, \bh^c \right)$ and 
$\bh ({\bf \bar{5}}) = \left( \bh, H^c \right).$ 
${\bf 5}, {\bf \bar{5}}$ are the representations of SU(5). 
$X(y), \bx(y)$ are the bulk fields in the ${\bf 24}$ dimensional 
representation under $SU(5)$.
\footnote{$X(y)$ and $\bx(y)$ are rescaled by $M_*^{-1/2}$, 
where $M_*$ is the Planck scale in five dimensional theory, 
since their mass dimension is 3/2 in five dimensions.} 
$\Sig$ is an usual SU(5) GUT adjoint Higgs field, 
which is assumed to be localized on the brane at $y=0$. 
The fifth dimensional coordinate is denoted by $y$. 
We assume that $X(y),\bx(y)$ depends on $y$, 
and $M,\bm$ do not. 
$\lm_{1\sim4}$ are dimesionless constants and 
$m_0$ is a bare mass parameter. 
This formulation of the action Eq.~(\ref{Higgsaction}) is useful 
because it is written by using the ${\cal N}=1$ superfield formalism 
and ${\cal N}=1$ SUSY is manifest \cite{AHHSW,NAHTGJW}.

F-flatness conditions of $X, \bx$ and $\Sig$ are 
\bea
\label{xf}
0 &=& \frac{\partial W}{\partial X} = 
\hc H - \frac{1}{5} 
{\rm tr}(\hc H) \nonumber \\ 
&& + \delta(y) \left\{ 2\lm_1 X \Sig
+ \lm_2 \Sig^2 - \frac{1}{5} 
\left( 2\lm_1 {\rm tr}(X\Sig) 
+ \lm_2 {\rm tr}(\Sig^2) \right) \right\}, \\
\label{bxf}
0 &=& \frac{\partial W}{\partial \bx} = 
\bhc \bh - \frac{1}{5} 
{\rm tr}(\bhc \bh) \nonumber \\
&&+ \delta(y) \left\{ 2\lm_3 \bx \Sig 
+ \lm_4 \Sig^2 - \frac{1}{5} 
\left( 2\lm_3 {\rm tr}(\bx\Sig) 
+ \lm_4 {\rm tr}(\Sig^2) \right) \right\}, \\
\label{sigf}
0 &=& \frac{\partial W}{\partial \Sig} = 
\delta(y) \left\{ \lm_1 X^2 + \lm_2 \bx^2 
+ 2\lm_3 X\Sig + 2\lm_4 \bx\Sig \right. \nonumber \\
&& \left. + m_0 \Sig - \frac{1}{5} 
\left( \lm_1 {\rm tr}(X^2) 
+ \lm_2 {\rm tr}(\bx^2) + 2\lm_3 {\rm tr} (X\Sig) 
+ 2\lm_4 {\rm tr}(\bx\Sig) \right) \right\},
\eea
where we omitted SU(5) indices for convenience. 
The trace part is proportional to the unit matrix. 
The solutions of Eqs.~(\ref{xf}) and (\ref{bxf}) are 
\bea
\label{xf1}
&& \hc H -\frac{1}{5} 
\tr (\hc H) = 0, \\
\label{xf2}
&& 2\lm_1 X(0) + \lm_3 \Sig = 0, \\
\label{bxf1}
&& \bhc \bh - \frac{1}{5} 
\tr (\bhc \bh) = 0, \\
\label{bxf2}
&& 2\lm_3 \bx(0) + \lm_4 \Sig = 0. 
\eea
It is remarkable that Eqs.~(\ref{xf2}) and (\ref{bxf2}) 
connect the vacuum expectation values (VEVs) $X(0)$ and $\bx(0)$ 
in the bulk with $\vev \Sig$ on the brane at $y=0$. 
As we will see later, 
$y$-independent masses of Higgs 
({\em i.e.} $X(0)+M$ and $\bx(0)+\bm$) determine 
the coordinates which Higgs wave functions are localized. 
These masses are different between the Higgs triplet 
and the Higgs doublet since $\langle X(0) \rangle$ 
and $\langle \bx(0) \rangle$ are 
proportional to $\vev\Sig$. 
Therefore, the splitting occurs naturally. 
Although a similar model has been considered 
in Ref.~\cite{maru,fatbrane}, 
they simply assumed that $\langle X(0) \rangle$ 
and $\langle \bx(0) \rangle$ take the form 
proportional to $\vev\Sig$. 
On the other hand, 
we {\em derived} this from the equations of motion.
\footnote{
The author would like to thank T.~Yanagida 
for suggesting that $\langle X(0) \rangle$ and 
$\langle \bx(0) \rangle$ should be derived from the potential. 
In Ref.~\cite{maru}, this point is not accomplished.} 
This is a crucial difference between Ref.~\cite{fatbrane} 
and this paper. 
Using Eqs.~(\ref{xf2}) and (\ref{bxf2}), 
Eq.~(\ref{sigf}) reproduces the stationary condition of 
the Higgs potential in the minimal SU(5) GUT, 
\be
0 = -\frac{3}{4} \left( \frac{\lm_3^2}{\lm_1} 
+ \frac{\lm_4^2}{\lm_2} \right) 
\left\{ \Sig^2 - \frac{1}{5} 
\tr (\Sig^2) \right\} 
+ m_0 \Sig. 
\ee
Furthermore, 
substituting $\vev\Sig = {\rm diag}(2, 2, 2, -3, -3)\sigma$, 
where $\sigma$ is a constant, we obtain 
\be
\frac{3}{2} \left( \frac{\lm_3^2}{\lm_1} 
+ \frac{\lm_4^2}{\lm_2} \right) \sigma + 2m_0 = 0. 
\ee
Expanding the five dimensional superfields 
$H, \hc, \bh$ and $\bhc$ by the mode functions as 
\bea
H(x,y) &=& \sum_n \phi_n(y) H_n(x), \\
\hc(x,y) &=& \sum_n \phic_n(y) \hc_n(x), \\
\bh(x,y) &=& \sum_n \bphi_n(y) \bh_n(x), \\
\bhc(x,y) &=& \sum_n \bphic_n(y) \bhc_n(x), 
\eea
where $x$ denotes the coordinate of the four dimensional space-time. 
The equations of motions 
for the zero mode wave functions of Higgs fields are 
\bea
(\pdy + X(y) + M)~\phi_0(y) &=& 0, \\
(- \pdy + X(y) + M )~\phic_0(y) &=& 0, \\
(\pdy + \bx(y) + \bm)~\bphi_0(y) &=& 0, \\
(-\pdy + \bx(y) + \bm)~\bphic_0(y) &=& 0. 
\eea

Let us assume for simplicity that $X(y) = X(0) + a^2 y$, 
$\bx(y) = \bx(0) + a^2 y$ in a small region 
of the point crossing zero. 
$a$ is a constant of mass dimension one. 
These mass functions generate Gaussian zero mode wave functions. 
The zero mode wave functions take the following form, 
\bea
\phi_0(y) &\sim& {\rm exp}\left\{ -\frac{a^2}{2} 
\left(y - \frac{X(0) + M}{a^2} \right)^2 \right\}, \\
\phic_0(y) &\sim& {\rm exp} \left\{\frac{a^2}{2} 
\left(y - \frac{X(0) + M}{a^2} \right)^2 \right\}, \\ 
\bphi_0(y) &\sim& {\rm exp} \left\{ -\frac{a^2}{2} 
\left(y - \frac{\bx(0) + \bm}{a^2} \right)^2 \right\}, \\
\bphic_0(y) &\sim& {\rm exp} \left\{ \frac{a^2}{2} 
\left(y - \frac{\bx(0) + \bm}{a^2} \right)^2 \right\}. 
\eea
Since the wave functions $\phic_0(y)$ and $\bphic_0(y)$ 
are not normalizable, its normalization constants must be zero. 
This result is consistent with Eqs.~(\ref{xf1}) and (\ref{bxf1}).

Now, we consider two cases 
which realize the doublet-triplet splitting. 
One is achieved through the bulk Higgs mass term \cite{maru,fatbrane}
and the other is achieved through 
the coupling of the singlet and the Higgs fields \cite{fatbrane}. 
First, we will show that 
the former case cannot incorporate 
the hierarchy of Yukawa couplings 
although the doublet-triplet splitting occur. 
The Higgs mass term in five dimensions is
\bea
&& \int d^4x dy \int d^2\theta M_* H(x,y) \bar{H}(x,y) \nonumber \\
&=& 
M_* \int dy \sqrt{\frac{a^2}{2\pi}}~{\rm exp} 
\left\{ -\frac{a^2}{2} \left(y - \frac{X(0) + M}{a^2} \right)^2 - 
\frac{a^2}{2} \left(y - \frac{\bx(0) + \bm}{a^2} \right)^2 \right\} 
\nonumber \\
&& \times \int d^4x \int d^2 \theta H_0(x) \bh_0(x). 
\eea
Higgs mass in four dimensions can be read by integrating out
degrees of freedom in the fifth dimension,
\be
\frac{M_*}{\sqrt{2}}~{\rm exp} 
\left\{- \frac{(X(0) + M - \bx(0) - \bm)^2}{4a^2} \right\} 
\int d^4x \int d^2 \theta H_0(x) \bh_0(x).
\ee
The masses of the Higgs triplets and the Higgs doublets are 
\bea
\label{colored}
&&M_3 \sim M_*~{\rm exp} 
\left[ - \frac{ \{2 (x-\barx) M_* + (m-\barm)M_*\}^2 }{4a^2} \right] 
\gsim M_{{\rm GUT}} \simeq 10^{16}~{\rm GeV}, \\
\label{weak}
&&M_2 \sim M_*~{\rm exp} 
\left[ - \frac{\{ -3 (x-\barx) M_* + (m-\barm)M_* \}^2}{4a^2} \right] 
\simeq M_W \simeq 10^2~{\rm GeV},
\eea
where $x, \barx$ and $m,\barm$ are defined as follows, 
\bea
X(0) &=& x~{\rm diag}(2,2,2,-3,-3)M_*, \\
\bx(0) &=& \barx~{\rm diag}(2,2,2,-3,-3)M_*, \\
M &=& m~M_*, \quad \bm = \barm~M_*.
\eea
We assumed here that the order of the VEV's 
of $X(0), \bx(0)$ and $M, \bm$ 
are around the five dimensional Planck scale $M_*$.

Before discussing the doublet-triplet splitting in detail, 
various scales in our model are summarized. 
There are three typical mass scales, {\em i.e.} 
the five dimensional Planck scale $M_*$, 
the wall thickness scale $L^{-1}$ 
which should be considered as the compactification scale 
and the inverse width of Gaussian zero modes $a^{-1}$. 
As explained in Ref. \cite{AS}, 
for the description  to make sense, 
the wall thickness $L$ should be larger than 
the inverse width of Gaussian zero modes $a^{-1}$. 
Furthermore, $a^{-1}$ should be larger than 
or equal to the five dimensional 
Planck length $M_*^{-1}$, 
\be
L^{-1} < a \le M_*. 
\ee
We take $L^{-1}$ to be $M_{GUT}$ in order to preserve 
the gauge coupling unification. 
The five dimensional Planck scale $M_*$ can be taken 
to be about $10^{17}$GeV or $10^{18}$ GeV from the above relation. 
Hereafter, $M_* \simeq 10^{18}$ GeV is taken for simplicity. 
In this case, 
the masses of the Higgs triplets (\ref{colored}) 
and the Higgs doublets (\ref{weak}) become 
\bea
&&{\rm exp} \left[ -\frac{\{ 2(\dx) + \dm \}^2}{4} \right] 
\gsim 10^{-2}, \\
&&{\rm exp} \left[ -\frac{\{-3(\dx) + \dm \}^2}{4} \right] 
\simeq 10^{-16}, 
\eea
where $a \simeq M_*$ is assumed for simplicity. 
These can be easily solved as 
\bea
\label{xdev}
-3.2844 \cdots \lsim &\dx& \lsim -1.5685 \cdots, \\
\label{mdev}
2.2793 \cdots \lsim &\dm& \lsim 7.4276  \cdots. 
\eea
This means that the doublet-triplet splitting is realized 
by ${\cal O}(1)$ tuning of parameters in contrast to 
an unnatural ${\cal O}(10^{14})$ fine-tuning of parameters 
in four dimensional case. 
As mentioned above, however, 
this case cannot reproduce the correct orders of 
magnitude of Yukawa couplings.
\footnote{There are several attempts to explain fermion mass 
hierarchy in the fat brane approach \cite{MS,KT}. 
The differences between Refs.~\cite{MS,KT} and this paper 
are the following. 
In Ref.~\cite{MS}, the wave functions of Higgs fields are flat 
in extra dimensions and non-supersymmetric case is considered. 
In Ref.~\cite{KT}, 
the wave functions of the matter are localized 
at the different points generation by generation.} 
In order to show this, we discuss $\dx \simeq -3$ 
and $\dm \simeq 3$ case as an example. 
In this case, 
the Higgs triplets $H_3, \bh_3$ are localized 
at $y \simeq (2x + m)M_*^{-1}, (2x + m + 3)M_*^{-1}$ 
and the Higgs doublets $H_2, \bh_2$ are localized at 
$y \simeq (-3x + m) M_*^{-1}, (-3x + m -12)M_*^{-1}$, respectively. 
Note that the relative distance between $H_2$ and $\bh_2$ is large. 
This is the problem. 
The left-handed quark superfield couples 
to both $H_2$ and $\bh_2$. 
In order to obtain ${\cal O}(1)$ top Yukawa coupling, 
the left-handed quark superfield 
of the third generation $Q_3$ and 
the right-handed quark superfield 
of the third generation $U^c_3$ 
must be localized around $H_2$. 
We will show that 
the correct order of magnitude of 
the bottom Yukawa coupling 
cannot be reproduced in this situation. 
The top Yukawa couplings in five dimensions are written by 
\bea
\label{yukawa}
&&\int d^4xdy \int d^2\theta 
\frac{Y_t}{\sqrt{M_*}}~
Q_3(x,y) U^c_3(x,y) H_2(x,y) + {\rm h.c.}  \\
&& \simeq \frac{Y_t}{\sqrt{M_*}} 
\left( \frac{2M_*^2}{\pi} \right)^{3/4} 
\int dy e^{-M_*^2 (y - y_{q_3})^2}e^{-M_*^2 (y - y_{u^c_3})^2}
e^{-M_*^2 (y-y_{h_2})^2} \nonumber \\
&&\times \int d^4x d^2\theta~
Q_{3,0}(x) U^c_{3,0}(x) H_{2,0}(x) + {\rm h.c},
\eea
where $Y_t$ is a top Yukawa coupling constant 
of order unity in five dimensions. 
We assumed that the zero mode wave functions of $Q_3, U^c_3$ 
and $H_2$ are also Gaussian and 
localized at $y \sim y_{q_3}, y_{u^c_3}$ 
and $y_{h_2}$, respectively. 
The effective top Yukawa coupling in four dimensions 
$y_t$ can be read as 
\be
\label{upyukawa}
y_t \sim Y_t~{\rm exp} \left[ -\frac{1}{3} M_*^2 
\left\{ (y_{q_3} - y_{u^c_3})^2 + (y_{q_3} - y_{h_2})^2 
+ (y_{u^c_3} - y_{h_2})^2 \right\} \right]. 
\ee
To be $y_t \sim {\cal O}(1)$, 
$y_{q_3} \simeq y_{u^c_3} \simeq y_{h_2}$.

On the other hand, 
the effective bottom Yukawa coupling in four dimensions are obtained 
by replacing $Y_t$ with $Y_b$ and 
$y_{u^c_3}, y_{h_2}$ with $y_{d^c_3}, y_{\bar{h}_2}$, 
\bea
y_b &\sim& Y_b~{\rm exp} \left[ -\frac{1}{3} M_*^2 
\left\{ (y_{q_3} - y_{d^c_3})^2 + (y_{q_3} - y_{\bar{h}_2})^2 
+ (y_{d^c_3} - y_{h_2})^2 \right\} \right], \\
&\sim& Y_b~{\rm exp} \left[ -\frac{1}{3} M_*^2 
\left\{ (y_{h_2} - y_{d^c_3})^2 + (y_{h_2} - y_{\bar{h}_2})^2 
+ (y_{d^c_3} - y_{\bar{h}_2})^2 \right\} \right], \\
&\lsim& Y_b~{\rm exp} \left[ -\frac{1}{3} M_*^2 
(y_{h_2} - y_{\bar{h}_2})^2 \right] 
\simeq {\rm exp} ( - 48) \simeq 10^{-21}, 
\eea
where $Y_b$ is a bottom Yukawa coupling constant 
of order unity, and 
$y_{q_3} \simeq y_{h_2}$ is used in the second line. 
Clearly, this is not realistic. 
Even if we take the other values satisfying 
Eqs.~(\ref{xdev}) and (\ref{mdev}) as $\dx$ and $\dm$, 
this result is not changed.

In order to improve this point, 
the Higgs triplets and the Higgs doublets 
are not only localized separately, 
but also the same multiplets have to be closely 
localized each other. 
Furthermore, 
the doublet-triplet splitting has to be realized 
by the overlap between the Higgs fields and the other bulk field, 
and by localizing the Higgs triplets close to this bulk field. 
This can be simply achieved by introducing 
the singlet field in the bulk \cite{fatbrane}.
\footnote{A similar coupling is considered, 
but the mechanisms to localize the bulk singlet 
are different between Ref.~\cite{fatbrane} and this paper.}

The action of the singlet sector is based on 
the ``shining" mechanism \cite{AHHSW} 
\be
S = \int d^4x dy \left[ \int d^4\theta (S^{\dagger}S + S^{c\dagger}S) 
+ \left\{ \int d^2\theta S^c (\partial_y + m_s) S 
- \delta(-y) \int d^2\theta JS + {\rm h.c.} \right\} \right], 
\ee
where $S$ is an SU(5) singlet superfield in the bulk, 
$S^c$ is its conjugated superfield, 
$J$ is a constant source and $m_s$ is a mass parameter. 
F-flatness conditions are 
\bea
\label{fs1}
0 &=& (\partial_y + m_s) S, \\
\label{fs2}
0 &=& (- \partial_y + m_s) S^c - J \delta(-y). 
\eea
%
The normalizable solutions of Eqs.~(\ref{fs1}) and (\ref{fs2}) are 
\bea
S &=& 0, \\
S^c &=& \theta(-y) J e^{m_s y}, 
\eea
where $\theta(y)$ is a step function for $y$. 

\noindent
The doublet-triplet splitting can be accomplished 
by introducing the following coupling.
\footnote{A similar coupling was considered in Ref. \cite{fatbrane}. 
They simply assumed the VEV of the singlet to be the GUT scale and 
does not specify the mechanism to generate the VEV.} 
\bea
&& \frac{1}{\sqrt{M_*}} \int d^4x dy \left\{ \int d^2 \theta 
S^c (x,y) H (x,y) \bh (x,y) + {\rm h.c.} \right\} \\
&& = \frac{1}{\sqrt{M_*}} \int dy S^c (y) \phi_0(y) \bphi_0(y) 
\int d^4x d^2 \theta H_0(x) \bh_0(x) + {\rm h.c.} \\
&& = \frac{M_*}{2\sqrt{2}} {\rm exp} 
\left[ - \frac{1}{2M_*^2} \left\{ (X(0) + M)^2 + (\bx(0) + \bm)^2 \right\} 
+ \frac{(X(0) + M + \bx(0) + \bm + m_s)^2}{4M_*^2} \right] \nonumber \\ 
&& \times \int d^4x d^2 \theta H_0(x) \bh_0(x) + {\rm h.c.}, 
\eea
where we assumed $J \simeq M_*^{3/2}, a \simeq M_*$.
\footnote{In order for the bulk Higgs mass term not to be allowed, 
we have to impose a symmetry, for example, an R-symmetry.} 
Therefore, the masses of the Higgs triplets and the Higgs doublets are 
\bea
\label{triplet2}
M_3 &\simeq& \frac{M_*}{2\sqrt{2}} {\rm exp} 
\left[ -\frac{1}{2} \{(2x + m)^2 + (2\bar{x} + \bar{m})^2 \} 
+ \frac{1}{4} (s + 2x + m + 2\bar{x} + \bar{m})^2 \right], \\
\label{doublet2}
M_2 &\simeq& \frac{M_*}{2\sqrt{2}} {\rm exp} 
\left[ -\frac{1}{2} \{(-3x + m)^2 + (-3\bar{x} + \bar{m})^2 \} 
+ \frac{1}{4} (s - 3x + m - 3\bar{x} + \bar{m})^2 \right], 
\eea
where we defined $m_s \equiv s M_*$.

If we consider the case that the Higgs triplets are localized 
at $y \simeq 0$, {\em i.e.} $2x + m \simeq 0$ and 
$2\bar{x} + \bar{m} \simeq 0$ 
for simplicity, the conditions of the doublet-triplet splitting are 
\bea
\label{triplet3}
M_3 &\sim& \frac{M_*}{2\sqrt{2}}~{\rm exp} (s^2/4) 
\gsim M_{{\rm GUT}} 
\sim 10^{16}~{\rm GeV}, \\
\label{doublet3}
M_2 &\sim& \frac{M_*}{2\sqrt{2}}~{\rm exp} 
\left[ - \frac{1}{2} \{(-5x)^2 + (-5 \bar{x})^2 \} 
+ \frac{1}{4} (s - 5x - 5\bar{x})^2 \right] \nonumber \\
&\simeq& M_W \simeq 10^2~{\rm GeV}. 
\eea
%
If we consider the case with $x = \bar{x}$, 
one of the solutions of Eq.~(\ref{doublet3}) 
is $x = \bar{x} \simeq 7$ 
and $s \simeq \sqrt{2}$. 
In this case, the mass of Higgs triplets is $M_3 \simeq 0.6M_*$. 
Eq.~(\ref{triplet3}) is satisfied since we are taking $M_*$ 
to be $10^{18}$ GeV. 
The Higgs triplets are localized at $y \simeq 0$, 
and the Higgs doublets are localized at $y \simeq -35M_*^{-1}$. 
The doublet-triplet splitting is realized 
by ${\cal O}$(1) tuning of the parameters 
in contrast to an unnatural fine-tuning 
in four dimensional case.

The next question is 
whether the following Yukawa coupling hierarchy 
can be obtained from the above setup
\footnote{We simply neglect the neutrino sector and the mixing angles 
since the detailed analysis of the fermion mass hiearchy 
and their mixing angles in our setup is not 
a main subject in this paper.}; 
\begin{center}
\begin{tabular}{lll}
$y_t \sim {\cal O}(1),$ & $\quad y_c \sim {\cal O}(10^{-2}),$ & 
$\quad y_u \sim {\cal O}(10^{-5}),$ \\
$y_b \sim {\cal O}(10^{-2}),$ & $\quad y_s \sim {\cal O}(10^{-4}),$ & 
$\quad y_d \sim {\cal O}(10^{-5}),$ \\
$y_\tau \sim {\cal O}(10^{-2}),$ & $\quad y_\mu \sim {\cal O}(10^{-4}),$ & 
$\quad y_e \sim {\cal O}(10^{-6}).$  
\end{tabular}
\end{center}
%
We would like to find from Eq.~(\ref{upyukawa}) 
the coordinates where the zero mode wave functions of the matter fields 
are localized and 
which induces the above hierarchy. 
We also take into account 
that the coefficients of the dimension five operators 
induced by the Planck scale physics 
$\frac{1}{M_P}QQQL$, that is $\frac{M_P}{M_*}~e^{-(M_*r)^2}$ where 
$r$ is the distance between the wave functions of 
quarks and the leptons, 
have to be less than $10^{-7}$ 
to keep the nucleon stable enough as required by 
experiments \cite{MK}. 
This constraints can be satisfied if $r \gsim (4\sim5)~M_*^{-1}$. 
The typical solution we found is 
\bea
&&y_{h_2} \simeq y_{\bar{h}_2} \simeq y_{q_3} \simeq y_{u^c_3} 
\sim -35 M_*^{-1},~y_{q_2} \simeq y_{u^c_2} \simeq y_{d^c_3} 
\simeq -37.6 M_*^{-1}, \\
&&y_{q_1} \simeq y_{u_1^c} \simeq y_{d_1^c} 
\simeq -39.1 M_*^{-1},~
y_{d_2^c} \simeq -38.7 M_*^{-1},~y_{l_3} 
\simeq -33 M_*^{-1}, \\
&&y_{e_3^c} \simeq -32.4 M_*^{-1},~y_{l_2} \simeq y_{e_2^c} 
\simeq -31.3 M_*^{-1},~y_{l_1} \simeq y_{e_1^c} 
\simeq -30.4 M_*^{-1}. 
\eea
We have checked that this configuration also satisfies 
the constraints for the coefficients of 
the dimension five operator $U^cU^cU^cE^c$.

In summary, 
we have discussed the doublet-triplet splitting problem 
in SUSY SU(5) GUT in five dimensions 
where the fifth dimension is non-compact. 
It was pointed out that 
an unnatural fine-tuning of parameters in order to obtain 
the light Higgs doublets is not required 
due to the exponential suppression of the overlap of the wave functions. 
We have found the explicit configuration of the Higgs and matter 
wave functions that realizes the doublet-triplet splitting, 
satisfies the constraints for the proton decay 
due to the dimension five operators 
induced by the Planck scale physics as well as 
by the Higgs triplet exchange 
and generates the correct orders of magnitude 
of Yukawa couplings. 
Furthermore, the gauge coupling unification is preserved 
because the inverse width of the fat brane $L^{-1}$ is the GUT scale.

There are some comments for our model to be more realistic. 
First, if we include the gravity we have to consider the warped extra 
dimension such as Randall and Sundrum \cite{RS2} model. 
In this case, the graviton localizes on the fat brane where 
the Standard Model fields are localized. 
Second, 
one may think that the localization of matter fields 
does not respect SU(5) symmetry. 
It is easy to improve this point. Since the localization point of 
the bulk fields are determined by the bulk mass parameters, 
these masses have only to respect SU(5) symmetry. 
In our model, 
this seems to be natural 
above the scale $\langle X \rangle \simeq \langle \bar{X} \rangle$ 
because SU(5) symmetry in five dimensions is unbroken.

Although the order of Yukawa couplings are explained, 
it is important to investigate 
whether the mixing angles can also be explained. 
Also, it is easy to incorpolate SUSY breaking in our setup 
(see Refs. \cite{AHHSW,KT,SS} due to the shining mechanism and 
Ref. \cite{MSSS} due to the coexistence of BPS domain walls.). 
It is very interesting to study the spectrum of the soft SUSY 
breaking terms in our setup, 
and investigate whether these spectrum satisfy 
the various experimental bounds. 
We leave these issues for future work.



\begin{center}
{\bf Acknowledgements}
\end{center}
The author would like to thank 
K.-I.~Izawa for a valuable discussion 
and T.~Yanagida for valuable discussions 
and a careful reading of the manuscript.
The author is supported 
by the Japan Society for the Promotion of Science 
for Young Scientists (No.08557).


\end{document}